\documentclass[11pt,a4paper]{llncs}
\usepackage{amsmath}
\usepackage{amssymb}
\usepackage{graphicx}
\usepackage{marvosym}
\usepackage{url}
\usepackage{fancyhdr}
\usepackage{cite}
\usepackage{hyperref}
\usepackage{float}
\usepackage{geometry}
\newcommand{\keywords}[1]{\par\addvspace\baselineskip\noindent\keywordname\enspace\ignorespaces#1}

\fancypagestyle{firstpage}{\fancyhf{}
\rfoot{\thepage}
}
\hypersetup{
    colorlinks=true,
    linkcolor=red,
    urlcolor=cyan,
    citecolor=blue,
    pdftitle={Integrating HW/SW Functionality for Flexible Wireless Radio},
}

\title{\LARGE{Integrating HW/SW Functionality for Flexible Wireless Radio}}
\author{\large{Alexander Strachan \and Nigel Topham}}
\institute{\large{School of Informatics, University of Edinburgh,\\ Edinburgh, Scotland, EH8 9AB\\ \email{alexander.strachan@ed.ac.uk} \quad \email{npt@ed.ac.uk}}}

\begin{document}
\maketitle
\thispagestyle{firstpage}

%%%%%%%%%%%%%%%%%%%%%%%%%%%%%%%%%%%%%%%%%%%%%%%%%%%%%%%%%%%%%%%%%%%%%%%%%%%%%%%%
\begin{abstract}
Current methods of implementing wireless radio typically take one of two forms;
either dedicated fixed-function hardware, or pure Software Defined Radio (SDR).
Fixed function hardware is efficient, but being specific to each radio standard it
lacks flexibility, whereas Software Defined Radio is highly flexible but requires
powerful processors to meet real-time performance constraints.

This paper presents a hybrid hardware/software approach that aims to combine
the flexibility of SDR with the efficiency of dedicated hardware solutions.

We evaluate this approach by simulating five variants of the IEEE 802.15.4 protocol,
commonly known as Zigbee, and demonstrate the range of performance and power consumption
characteristics for different accelerator and software configurations.
Across the spectrum of configurations we see power consumption varies from 8\% to 38\%
of a dedicated hardware implementation, and show how the hybrid approach allows
a new modulation standard to be retrofitted to an existing design, with only a modest
increase in power consumption.

\end{abstract}

%%%%%%%%%%%%%%%%%%%%%%%%%%%%%%%%%%%%%%%%%%%%%%%%%%%%%%%%%%%%%%%%%%%%%%%%%%%%%%%%
\section{Introduction}
Wireless radio is a pervasive technology, which relies heavily on real-time
data streaming and digital signal processors. However, architectures for wireless
radio tend to be specialized, fixed-function devices designed to handle high data
rates and complex mathematical processing.

Current embedded wireless radio designs operate across a spectrum and can
be summarized at three extremes;
fixed-function radio PHYs where all processing is done in dedicated hardware;
software defined radio where all processing is done in software;
and offload engines where a proportion of processing is performed in software before
passing it to fixed-function hardware.
The above methods all share the trait that once data is passed to fixed function
hardware, that data is never seen again by the CPU.
This has disadvantages; if the hardware does not perfectly fit the
required algorithm it cannot be used, and if it cannot be disabled, then
processing would need to be moved to software, decreasing efficiency.

The system architecture presented in this paper splits a hardware accelerator into a pipeline
of computational blocks, each representing one stage of the radio processing pipeline,
but allows the CPU to optionally intercept and manipulate data as it moves between
pipeline stages.
This provides flexibility for each part of the pipeline to be implemented in
hardware or software, and for additional pipeline stages to be inserted as
software components, as required by the modulation scheme.
The IEEE 802.15.4\cite{zigbee} standard, commonly referred to as Zigbee, is used
in this paper as an example of a simple, industry standard radio specification
used in IoT applications such as Thread\cite{thread}, that contains several
different modulation schemes.

%%%%%%%%%%%%%%%%%%%%%%%%%%%%%%%%%%%%%%%%%%%%%%%%%%%%%%%%%%%%%%%%%%%%%%%%%%%%%%%%
\subsection{Research Aims and Challenges}
The primary motivation of this research is to explore how radio-processing systems
can be made more flexible by sharing the workload of PHY processing between
accelerator hardware and a general-purpose processor.

Key questions include;
Can radio hardware designs be fixed/enhanced later on using software/CPUs?
How does performance vary across the spectrum between pure hardware and software?
What is the performance/power consumption impact on the system?
Is there a cutoff point beyond which software cannot keep up?
What phases of radio processing demand acceleration, and what phases can be
adequately handled by software?
How should data be streamed between hardware accelerator and software running on a CPU?

To answer these questions requires a simulation system capable of exploring the
design space of hybrid software-hardware architecture solutions.
This simulation system should accurately model real-world CPU and radio designs, by
generating bit-accurate radio data streams, providing representative data on the
performance of the system, and demonstrating the relative impact on power
consumption, when hardware functions are handled in software.

%%%%%%%%%%%%%%%%%%%%%%%%%%%%%%%%%%%%%%%%%%%%%%%%%%%%%%%%%%%%%%%%%%%%%%%%%%%%%%%%
\section{Accelerator Architecture}

\begin{figure}
    \centering
    \includegraphics[width=0.5\linewidth]{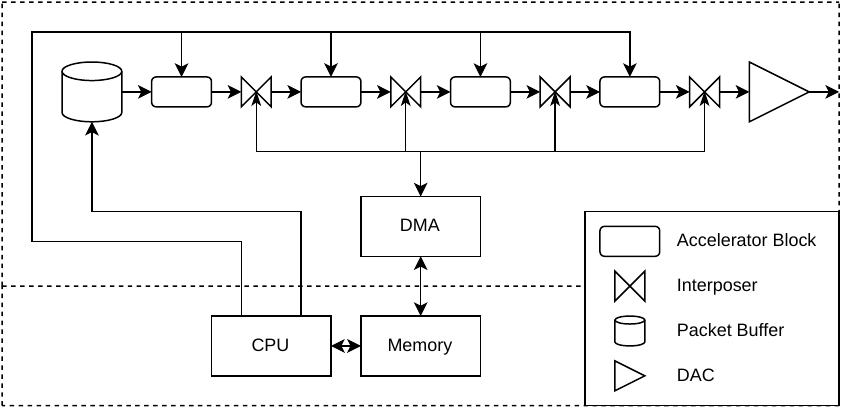}
    \caption{Overall System Diagram showing accelerator and interposer connections to CPU and memory hierarchy.}
    \label{fig:system}
\end{figure}

Figure~\ref{fig:system} shows the overall architecture of a system combining
a general-purpose CPU with a radio-processing accelerator pipeline. This shows only
the transmitter side, although the same approach can be applied to the receiver.
The design comprises a packet buffer, where the data packets to be modulated are stored,
a number of partially reconfigurable accelerator blocks that perform the radio
signal processing, and a Digital to Analog Converter (DAC).
However, such a design, while partially reconfigurable, is not flexible enough to
support a hybrid system, that allows software to assist the processing.
To allow this, devices called interposers are inserted between each stage of
the accelerator pipeline.
These interposers interface to the CPU via a DMA device, allowing the CPU to
extract or insert data at any point in the accelerator pipeline. The CPU is
then able to create flexibility by adding or replacing stages
in the pipeline using software components.

%%%%%%%%%%%%%%%%%%%%%%%%%%%%%%%%%%%%%%%%%%%%%%%%%%%%%%%%%%%%%%%%%%%%%%%%%%%%%%%%
\subsection{Radio Accelerator Pipeline Design}
The IEEE 802.15.4 standard~\cite{zigbee} specifies many different Physical Layer
(PHY) designs intended for different applications.
For this work, six variations of this standard were chosen that are similar in
complexity and processing, and share mathematically common features.
These can be seen in Table~\ref{tab:std}.
\begin{table*}
\caption{IEEE 802.15.4 modulation types, data/symbol/sample rates (at four samples per symbol), for various specified frequency bands.}
\label{tab:std}
\centering
\setlength\tabcolsep{5pt}
\begin{tabular}{|l|c|c|r|r|r|}
\hline
No. & Frequency Band & Modulation & Data Rate & Symbol Rate  & Sample Rate\footnotemark \\ \hline
1   & 2450 MHz       & OQPSK      &  31250 B/s &  62500 sym/s & 4 MHz       \\
2   & 915 MHz        & OQPSK      &  31250 B/s &  62500 sym/s & 2 MHz       \\
3   & 780 MHz        & OQPSK      &  31250 B/s &  62500 sym/s & 2 MHz       \\
4   & 868 MHz        & BPSK       &   2500 B/s &  20000 sym/s & 1.2 MHz     \\
5   & 915 MHz        & BPSK       &   5000 B/s &  40000 sym/s & 2.4 MHz     \\
6   & 920.8–928 MHz  & GFSK       & 12500 B/s  & 100000 sym/s & 400 kHz     \\ \hline
\end{tabular}
\end{table*}

\begin{figure}
    \centering
    \includegraphics[width=0.6\linewidth]{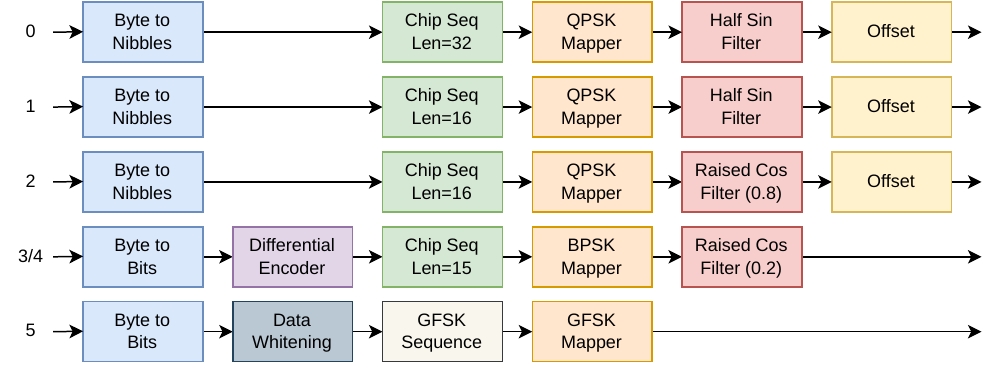}
    \caption{Radio Accelerator Pipelines for The Chosen 802.15.4 Standards.}
    \label{fig:blockpipes}
\end{figure}

\begin{figure*}
    \centering
    \includegraphics[width=0.8\linewidth]{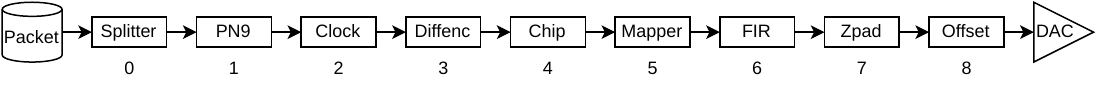}
    \caption{Unified Reconfigurable Radio Accelerator Pipeline.}
    \label{fig:blockpipe}
\end{figure*}

In Figure~\ref{fig:blockpipes}, block diagrams for each of the six standards
are shown. It can be seen that these six standards share common operations or
have operations that are a superset or a subset of operations in other
standards.
Therefore, a unified set of partially reconfigurable blocks was designed and
then combined to form a single processing pipeline. This pipeline can be seen
in Figure~\ref{fig:blockpipe} and can be reconfigured to support any one of the
six standards, either by disabling blocks that are not required, or by
reconfiguring blocks to support a given radio standard.
Descriptions explaining the operation of each block are shown in Table~\ref{tab:blockdesc}.

\begin{table}[h]
\caption{Unified Accelerator Block Descriptions.}
\label{tab:blockdesc}
\centering
\begin{tabular}{|r|l|}
\hline
Block    & Description                                     \\ \hline
Splitter & Splits bytes into symbols (nibbles or bits).    \\
PN9      & XORs a data whitening sequence with symbols.    \\
Clock    & Generates (counter)clockwise series of symbols. \\
Diffenc  & Differential encoder.                           \\
Chip     & Maps symbols to orthogonal chip sequences.      \\
Mapper   & Maps symbols to complex IQ samples.             \\
FIR      & 41-tap digital filter, used to shape samples.   \\
Zpad     & Inserts N zeros every M samples.                \\
Offset   & Delays imaginary (Q) component by N samples.    \\ \hline
\end{tabular}
\end{table}
\footnotetext{Using four samples per symbol}

%%%%%%%%%%%%%%%%%%%%%%%%%%%%%%%%%%%%%%%%%%%%%%%%%%%%%%%%%%%%%%%%%%%%%%%%%%%%%%%%
\subsection{Interposer Design}
\begin{figure}
    \centering
    \includegraphics[width=0.4\linewidth]{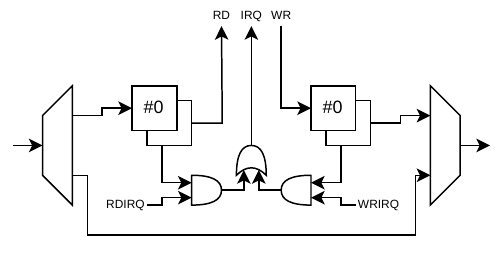}
    \caption{Block Diagram Showing Interposer Design.}
    \label{fig:interposer}
\end{figure}

To support a hybrid design, an interposer was designed which is inserted between
blocks in the accelerator pipeline.
A diagram showing the interposer design can be seen in Figure~\ref{fig:interposer}.
It consists of; a set of double buffers, one set for each direction, allowing data
to be filled and drained from the accelerator pipeline simultaneously with data
transfer to the CPU;
a pair of (de)multiplexers, allowing the interposer to be enabled or disabled, and
logic to interrupt the CPU when there is a certain amount of data to be read or
written to the interposer buffer.
Data transfer to and from the interposers utilizes DMA, to handle the high data
rates required.

\begin{figure}
    \centering
    \includegraphics[width=0.4\linewidth]{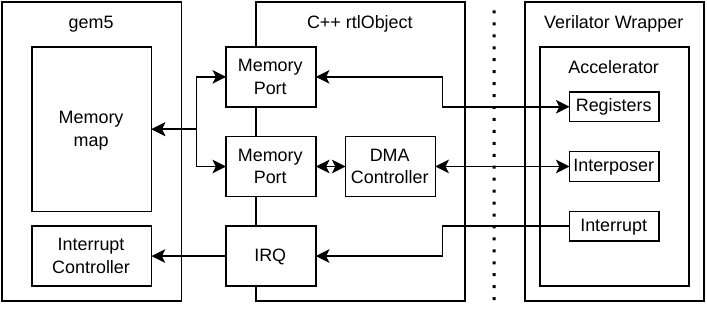}
    \caption{Gem5 Simulation Architecture Overview.}
    \label{fig:gem5}
\end{figure}

Initially, the accelerator was connected to the CPU as an I/O device,
but this led to unacceptable performance because the accelerator interface
registers were treated as uncachable locations. This led to significant cache
coherency overheads when moving data from cached data structures created by
the CPU to the uncached accelerator I/O registers (and {\em vice versa}).

The introduction of DMA allowed burst transfers to be used to transfer data to and from the
accelerator, but this still had unacceptable performance, as the mapped DMA memory
still needed to be marked as uncachable.
By using manual cache flush/invalidate operations on the DMA memory, these costs
were reduced significantly, allowing slow memory transfer operations to be
performed in longer bursts.
This system allows a general purpose CPU to intervene and manipulate data from
the accelerator pipeline, without unduly affecting the real-time
characteristics of the system.
Additionally, a ring buffer was added between the last block and the DAC to
smooth out any non-determinism in the CPU response time.

%%%%%%%%%%%%%%%%%%%%%%%%%%%%%%%%%%%%%%%%%%%%%%%%%%%%%%%%%%%%%%%%%%%%%%%%%%%%%%%%
\subsection{Software Design}
A software implementation of the Zigbee radio standards was written in \texttt{C++}
to run under Linux as a userspace application.
The Linux UIO system~\cite{UIO} is used to allow a userspace application to
respond to CPU interrupts by blocking on a \texttt{read(2)} system call,
allowing the CPU to sleep when no data transfer is required.
Additionally, the u-dma-buf kernel module~\cite{udmabuf} allows a userspace
application to \texttt{mmap(2)} memory suitable for DMA operations. It also allows
the program to manage cache coherency on this memory, allowing the userspace
application to invalidate and flush the CPU cache.

%%%%%%%%%%%%%%%%%%%%%%%%%%%%%%%%%%%%%%%%%%%%%%%%%%%%%%%%%%%%%%%%%%%%%%%%%%%%%%%%
\subsubsection{Operation}
First, the CPU fills the packet buffer, configures each pipeline stage via
memory mapped I/O and starts the accelerator. If the interposers are not in
operation, there is nothing more for the CPU to do.
The CPU then reads the accelerator status registers to determine if the enabled
interposer is ready for reads or writes.
If no interposer is ready, the \texttt{read(2)} system call is used on the
UIO device, which causes the process to block, until the next interrupt occurs.
When the CPU is unblocked, it again checks to see if the accelerator is
waiting on a read or write.
A read is accepted if the CPU currently has no pending data to write, then the
size of the data is checked and the cache backing the DMA buffer is invalidated,
by writing to a file exposed by the u-dma-buf module. If data is to be processed
by software, this is done now, leaving an output buffer ready to be written when
the interposer is ready.
A write is accepted if the CPU has pending data to write. If the data to be
written is larger than the currently set buffer size, it is broken up into chunks
and the size of this data is then written to the accelerator.
The data is then written to the mapped DMA buffer, and the cache is flushed,
by writing to a file exposed by u-dma-buf. Then the DMA transfer is
started, and various counters are updated, depending on how much data is left to
be transferred.
This whole process repeats, until a packet is read with the ``last''
flag set, indicating that this is the last data packet to be transferred. Once
the last data packet is processed, the process sleeps until another interrupt
occurs which signifies the accelerator as finished.

%%%%%%%%%%%%%%%%%%%%%%%%%%%%%%%%%%%%%%%%%%%%%%%%%%%%%%%%%%%%%%%%%%%%%%%%%%%%%%%%
\section{Results}
To obtain experimental results for the above design, a simulator was designed
based on gem5+rtl~\cite{gem5rtl}. This approach allows cycle
accurate hardware RTL designs to be simulated in conjunction with a simulated
CPU model by integrating gem5~\cite{gem5} and Verilator~\cite{Verilator}.
gem5 is used here as a generic model of an out-of-order CPU and its configuration
is not changed between simulation runs.
Therefore, any potential inaccuracies in gem5's micro-architectural model will
apply equally to all results.

A block diagram of the simulator architecture can be seen in Figure~\ref{fig:gem5}.
It consists of two modules, the accelerator design written in SystemVerilog,
which is connected to a gem5 device written in C++.
The Verilog module contains the accelerator blocks and interposers which are
connected together via AXI-Stream~\cite{AXIStream} interfaces.
The C++ module forwards memory and interrupt requests to and from the Verilog
simulation and the CPU.
As the RTL system does not provide a way to directly connect Verilog code to the
gem5 memory hierarchy, the C++ module contains a simple DMA controller.
The accelerator peripheral is connected to the CPU via the off-chip peripheral
bus.

To evaluate the design, tests were constructed that simulated the transmission
of a number of packets using the hybrid CPU/accelerator architecture.
These tests allowed any contiguous segment of the nine accelerator
blocks to be replaced with a software equivalent, enabling the behavior of
the system to be analyzed when different parts of the system are replaced
with software.
After each accelerator run, the generated IQ data was written to a file to
allow the accelerator output to be checked for correctness.
This output was compared against reference models in MATLAB~\cite{MATLAB} and
GNURadio~\cite{GNURadiozigbee}.

%%%%%%%%%%%%%%%%%%%%%%%%%%%%%%%%%%%%%%%%%%%%%%%%%%%%%%%%%%%%%%%%%%%%%%%%%%%%%%%%

\subsection{CPU Power Consumption}
To evaluate the impact on power consumption of using software to assist a
hardware accelerator, simulations were run with each block in the hardware
pipeline in turn being replaced with an equivalent software module running on the CPU.
In these tests the CPU is idle when gem5 reports the CPU as clock-gated,
and therefore the proportion of all cycles that are clock-gated is used as
an indicator of the relative dynamic power consumption of the CPU.
These experiments focused on the OQPSK modulation scheme operating at 2450MHz,
as this has the highest data rate and thus presents a worse case scenario.

\begin{figure}
    \centering
    \includegraphics[width=0.4\linewidth]{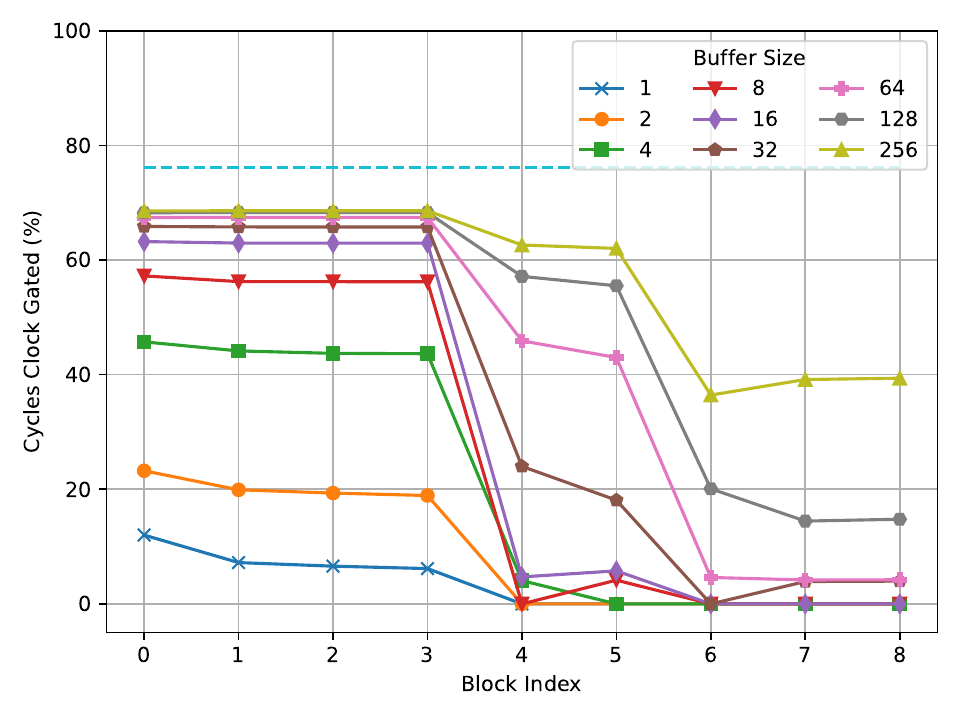}
    \caption{Cycles Clock Gated Across Each Block Replaced with Software for Various Buffer Sizes}
    \label{fig:sw}
\end{figure}

This test was performed for a range of different buffer
sizes and results were compared against a full hardware implementation,
as shown in Figure~\ref{fig:sw}.
It can be seen that for blocks appearing earlier in the pipeline, and with a large enough
buffer size, the performance of this architecture can reach to within $10\%$ of
a full hardware implementation. This gap widens when blocks from further down the accelerator
pipeline are replaced by software.
This test was repeated with the software processing disabled and
the hardware left enabled, to measure how much overhead the
software signal processing has over the cost of data movement.
It was seen that the software processing overhead was between $\pm 6\%$.
This shows that the primary cost of this system is the movement of data, not
the complexity of the signal processing.

%%%%%%%%%%%%%%%%%%%%%%%%%%%%%%%%%%%%%%%%%%%%%%%%%%%%%%%%%%%%%%%%%%%%%%%%%%%%%%%%

\begin{figure}
    \centering
    \includegraphics[width=0.4\linewidth]{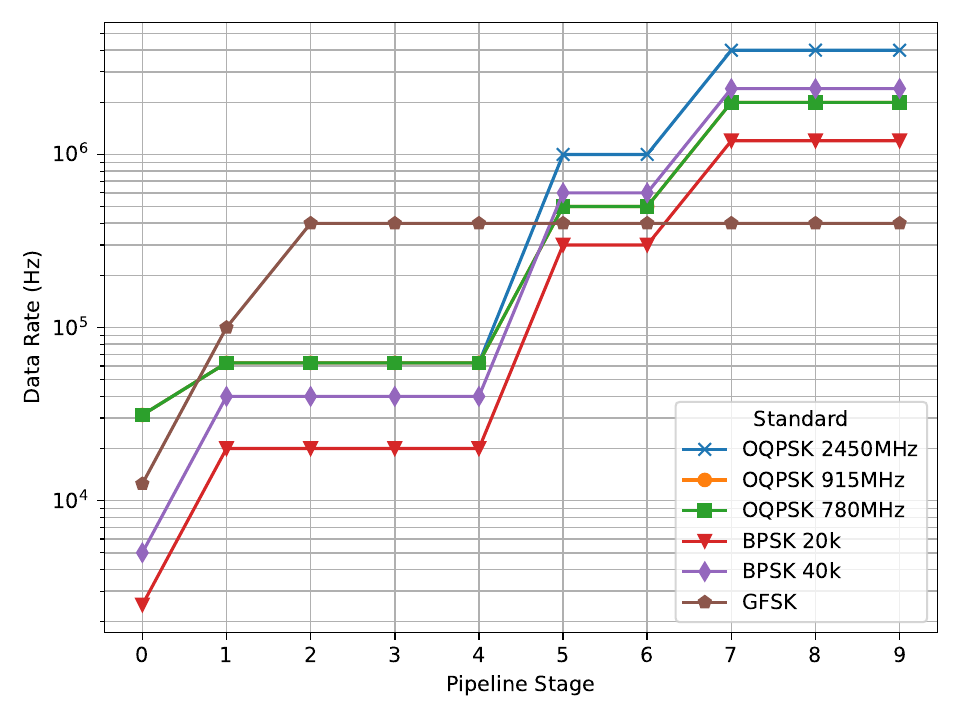}
    \caption{Data Rate Across Accelerator Pipeline per Radio Standard}
    \label{fig:rate}
\end{figure}

\begin{figure}
    \centering
    \includegraphics[width=0.4\linewidth]{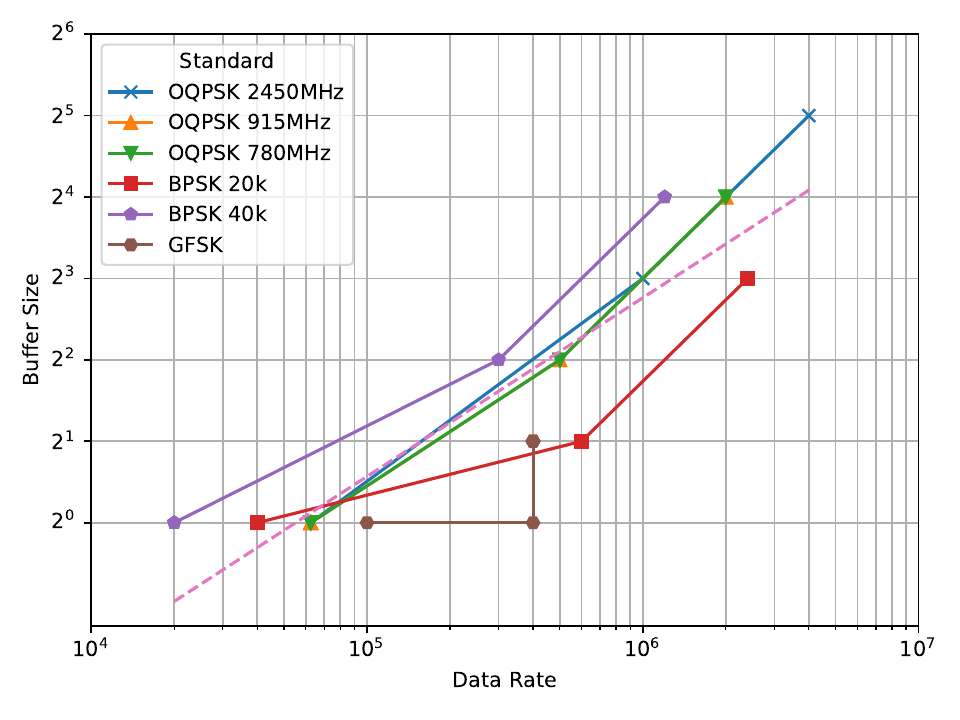}
    \caption{Minimum Buffer Size for Data Rate Relation}
    \label{fig:ratebuf}
\end{figure}

The above tests show the primary indicator of performance is the data rate that
a block is operating at. Figure~\ref{fig:rate} shows the data rate increase
across the pipeline for each standard, it can be seen that this plots shape is
similar to that of Figure~\ref{fig:sw}.
However, it was noted that certain combinations of block index and buffer size
would cause the DAC to underrun, meaning that this combination is unsuitable
for real-time applications. Figure~\ref{fig:ratebuf} shows the minimum buffer size
to not underrun, against data rate. Noting the log-log scale, a power relation
can be seen, $Size \approx k\cdot rate^m$ where $m\approx 0.66, k\approx 0.0007$.
This is a minimum bound, in a real system, a higher buffer size may be required.

%%%%%%%%%%%%%%%%%%%%%%%%%%%%%%%%%%%%%%%%%%%%%%%%%%%%%%%%%%%%%%%%%%%%%%%%%%%%%%%%

\begin{table*}[t]
\caption{Block Usage per Modulation}
\centering
\begin{tabular}{|l|ccccccccc|}
\hline
           & Splitter & PN9    & Clock  & Diffenc & chip   & mapper & FIR    & zpad   & offset \\ \hline
OQPSK      & \checkmark   & ---    & ---    & ---     & \checkmark & \checkmark & \checkmark & \checkmark & \checkmark \\
BPSK       & \checkmark   & ---    & ---    & \checkmark  & \checkmark & \checkmark & \checkmark & ---    & ---    \\
GFSK       & \checkmark   & \checkmark & \checkmark & ---     & ---    & \checkmark & ---    & ---    & ---    \\ \hline
\end{tabular}
\label{tab:blkuse}
\end{table*}

\begin{figure}
    \centering
    \includegraphics[width=0.4\linewidth]{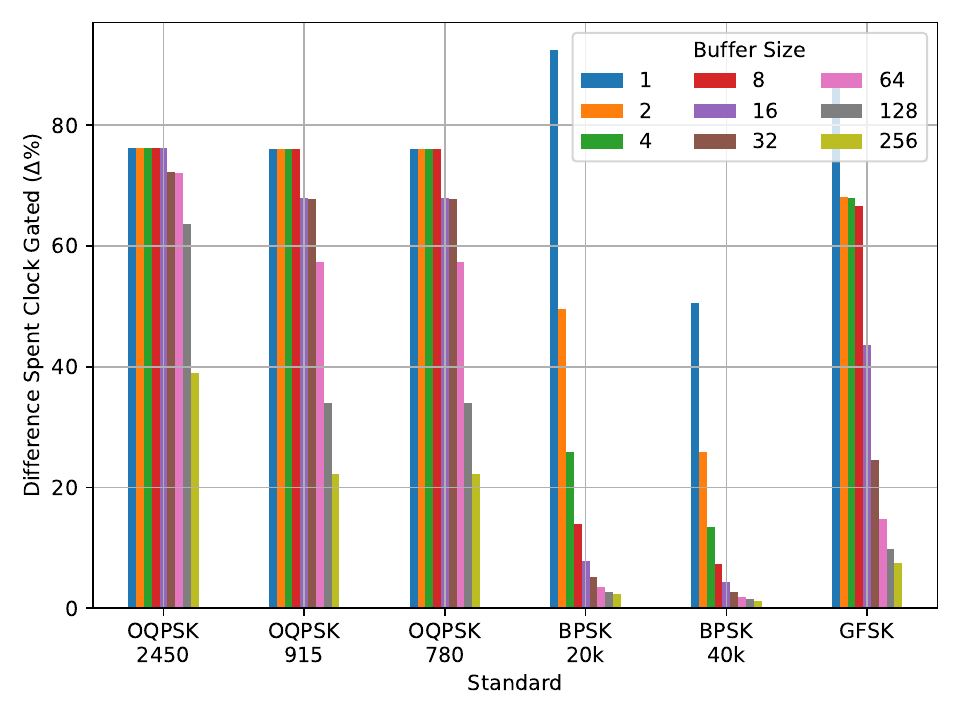}
    \caption{Cycles Spent Clock Gated Across Each Standard with Unique Blocks Removed for Various Buffer Sizes}
    \label{fig:rep}
\end{figure}

It can be seen in Table~\ref{tab:blkuse}, that certain blocks in the accelerator
pipeline are only used by one standard. This opens up the possibility of
simulating a system where these blocks were not part of the
accelerator when it was manufactured.
These ``missing'' blocks could then be added using software instead, allowing
a radio standard to be added ``after the fact''.
Therefore, an experiment was run where these blocks were disabled and replaced with software
and these results were compared with a full hardware
implementation of that standard, for a range of buffer sizes.
It can be seen in Figure~\ref{fig:rep} that in the best case (BPSK), with
a large enough buffer size, power consumption can be within $3\%$ of a full
hardware implementation, but in the worst case (OQPSK 2450) power consumption
is $40\%$ worse than hardware.
This difference is due to OQPSKs unique blocks occurring later in the pipeline, and
thus they operate at a higher data rate.

%%%%%%%%%%%%%%%%%%%%%%%%%%%%%%%%%%%%%%%%%%%%%%%%%%%%%%%%%%%%%%%%%%%%%%%%%%%%%%%%
\subsection{CPU Behaviour}
In the prior section it was seen that the performance impact of this
architecture is strongly linked with the movement of data to and from the
accelerator.
To further explore the CPU behaviour during these periods, the test program was
modified to add marks at each phase of the programs execution.
These phases are: \texttt{Init} where the system is configured; \texttt{Loop}
where decisions are made on what operation to perform; \texttt{IRQ} where if the
CPU has no work to do, it sleeps; \texttt{Read} is where data is read from the
accelerator; \texttt{DSP} where data processing is performed; \texttt{Write}
is where data is written to the accelerator; and \texttt{End} where the CPU is
finished processing and is sleeping waiting for the accelerator to complete.
The tests were run for each standard at a buffer size of 256.

\begin{figure}
    \centering
    \includegraphics[width=0.4\linewidth]{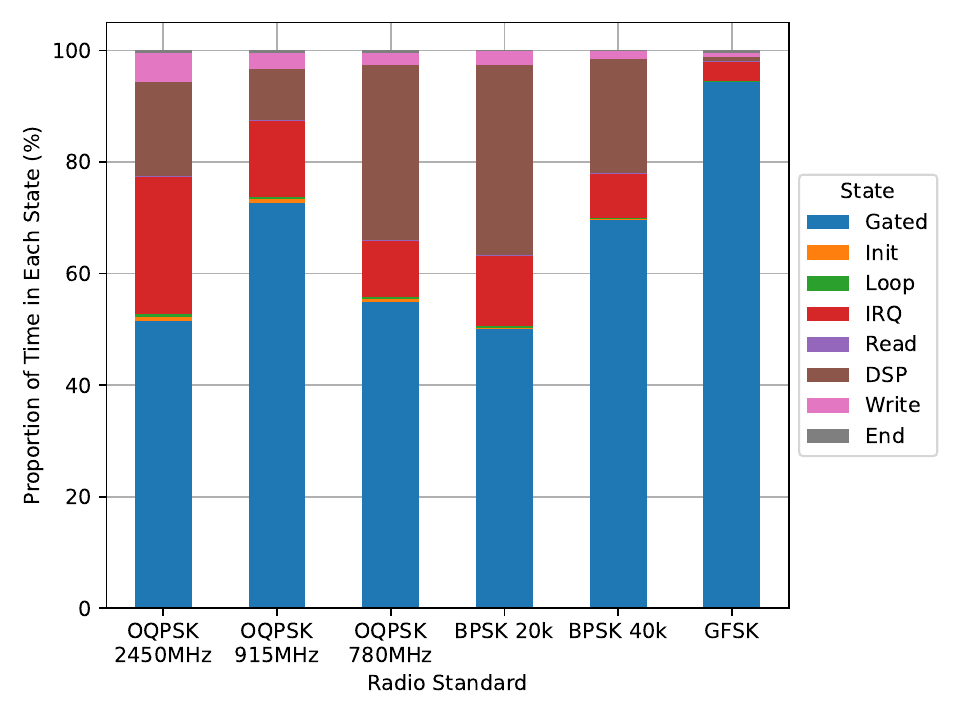}
    \caption{Proportion of CPU Cycles Spent in Each State per Modulation Type}
    \label{fig:statetime}
\end{figure}
\begin{figure}
    \centering
    \includegraphics[width=0.4\linewidth]{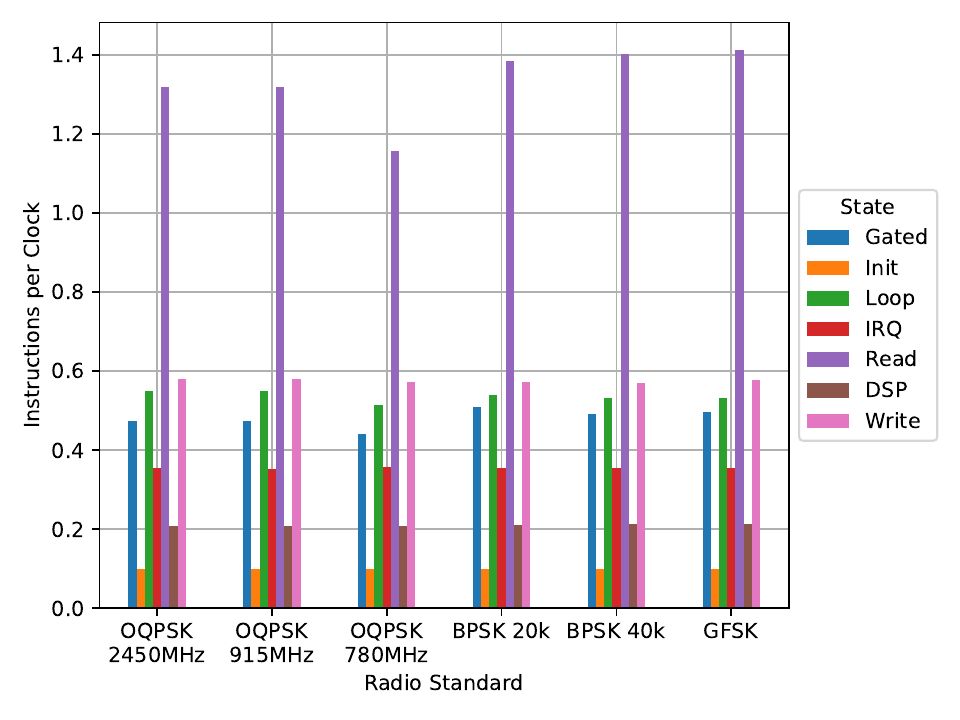}
    \caption{Instructions per Clock in Each State per Modulation Type}
    \label{fig:ipc}
\end{figure}

In Figure~\ref{fig:statetime} the time spent in each phase is shown. It can be seen
that the CPU spends the majority of its time clock gated, leading to reduced
power consumption.
The time the CPU spends active is primarily spent in the DSP, IRQ, Read and Write
stages.
Across all standards the average IPC is $0.25$, showing poor instruction level
parallelism.
In Figure~\ref{fig:ipc}, the IPC in each phase is shown, showing that while the
DSP takes up the majority of the time, it is by far the most efficient phase.
Whereby the other phases show IPC values well below $0.6$.
These results show that the time spent in the IRQ, read and write stages are
inefficient enough to bring down the overall IPC by a factor of $5$, despite them
taking less overall time and being less mathematically complex
than the signal processing.

\begin{figure}[H]
    \centering
    \includegraphics[width=0.4\linewidth]{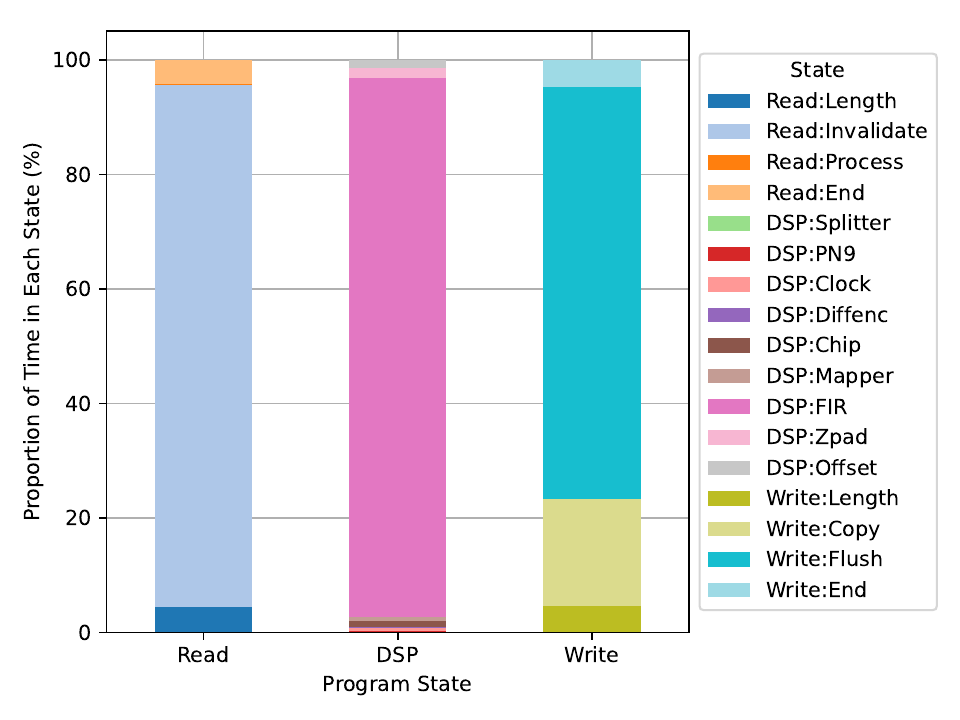}
    \caption{Breakdown of CPU Cycles Spent in Read DSP and Write States}
    \label{fig:readdspwrite}
\end{figure}

To investigate potential causes of the low IPC, the cycles spent and state of
the CPU pipeline was measured for each phase of the program.
In Figure~\ref{fig:readdspwrite} a breakdown of the cycles spent in the
Read Write and DSP phases can be seen, averaged across all standards. It can be
seen that the most computationally intensive signal processing block is the FIR
filter and that most of the time spent reading and writing is on cache management
operations, relating to the setup of the DMA unit.

\begin{figure}
    \centering
    \includegraphics[width=0.4\linewidth]{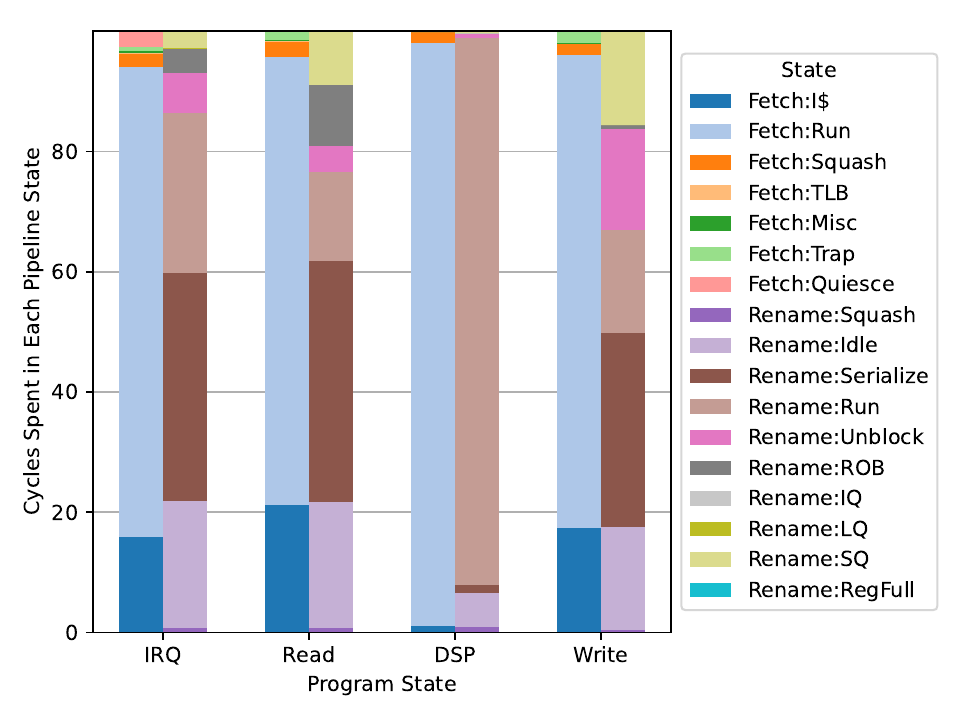}
    \caption{Breakdown of Cycles Spent in Fetch and Rename Stages in Read DSP and Write States}
    \label{fig:pipeline}
\end{figure}

In Figure~\ref{fig:pipeline} a breakdown of the CPU fetch and rename pipeline
states is shown for the IRQ, Read, Write and DSP stages, averaged across each
standard.
The IRQ state spends a significant amount of time stalled in the fetch state,
due to instruction cache misses, caused by the system call overhead, as the
kernel manages the interrupts. On average the interrupt latency is $7494$
cycles.
The read and write stages spend a similar proportion of time also stalled on the
instruction cache, also due to the cache management operations requiring system
calls.
Additionally, the cache management operations, and memory reads/writes stall the
CPU in the rename stage, due to the need to serialize data and the load/store queues
and ROB becoming full.
Therefore, it can be seen that the cause of the low IPC is the large overhead of
moving memory between the CPU and the accelerator and management of the CPU caches.

%%%%%%%%%%%%%%%%%%%%%%%%%%%%%%%%%%%%%%%%%%%%%%%%%%%%%%%%%%%%%%%%%%%%%%%%%%%%%%%%
\section{Conclusion}
In conclusion, this work shows that software can be used to enhance a hardware
accelerator pipeline, and can work effectively with a real-world modulation
standard, in real-time.
A relation was discovered that can be used to calculate the minimum buffer size
required for software to intervene at a given data rate.
The extra power consumed by this design varies considerably depending on
the data rate at the which the CPU is required to intervene.
However, is it hoped that this overhead can be reduced in future designs.
Finally, a co-simulation approach was used that allowed cycle-accurate
simulation of the hardware accelerator, so it could be integrated in a SoC
design with minimal effort.

However, communication between the CPU and interposers has a large overhead due
to the interposer and CPU having opposing views on memory layout.
The CPU requires a cache hierarchy, where data is stored at various non-local
levels to hide DRAM latency; while the accelerator requires contiguous data in
a single location to take advantage of DMA burst transfers.
This results in a significant amount of time being spent of cache management
operations, that cause the CPU to stall, due to lack of space in the ROB and
load/store queues. Cache management and interrupt handling also require system
calls that cause high instruction cache miss rates, also stalling the CPU.

A limitation with this approach, is that certain radio standards may have areas
in their processing pipeline which operate at too high a data rate for a CPU to 
replace.
Additionally, a potential drawback of this system is that the overall performance 
of the SoC would be reduced because one of the CPU cores would be tied up performing 
signal processing.
If this system can be used to add or enhance a standard, it will come at the
cost of increased power consumption, with the tradeoff being that, the SoC will 
not need to be redesigned or replaced, reducing development time and waste.
A more detailed comparison of this system, against existing hardware, is planned
using a future improved version of the system described in this paper.

This overhead results in the CPU reporting low IPC values, due to poor
instruction-level parallelism, leading to more clock cycles spent active,
increasing power consumption.
While the proportion of CPU time spent on digital signal processing is large,
the IPC value during this period is high, showing that the CPU is operating
efficiently during this time.
However, it is unclear if the low efficiency of the cache management operations
are due to the operation itself, or the cost of the system calls required to
perform these operations.

%%%%%%%%%%%%%%%%%%%%%%%%%%%%%%%%%%%%%%%%%%%%%%%%%%%%%%%%%%%%%%%%%%%%%%%%%%%%%%%%

\subsection{Related Work}
\subsubsection{Research}
This work builds upon the gem5+RTL framework~\cite{gem5rtl}, which allows
collection of detailed CPU behavior information, while using an RTL model for the
accelerator design, combining the advantages of both types of simulation.

Similar work in reconfigurable radios for Zigbee have been considered
in~\cite{zreconfig1,zreconfig2} but neither of these papers combine
similar blocks for different modulation types.
In papers~\cite{reconfig1,reconfig2,reconfig3} this combining
of similar processing blocks is performed for the Wi-Fi and 3G protocols.

The RFNoC~\cite{rfnoc} project looks into taking advantage of the FPGAs that are
present in many SDR interfaces, in the GNURadio framework, to alleviate the CPU from
the most demanding processing. However, the use of high-level software and FPGAs preclude
use of this in embedded systems.

Additionally, while pure software solutions may work for simpler standards such
as Zigbee, other more complicated standards like Wi-Fi require parts of the
PHY and MAC to have hardware support to meet timing
deadlines~\cite{wifi1,wifi2,wifi3,wifi4,wifi5,wifi6}.
This shows that some sort of hardware acceleration can be required, even if power
consumption is not taken into account.

However, none of this work considers a combined hybrid approach of these ideas,
placing the CPU ``in the loop'' between accelerator blocks, in a way that could
work in an embedded system context.

%%%%%%%%%%%%%%%%%%%%%%%%%%%%%%%%%%%%%%%%%%%%%%%%%%%%%%%%%%%%%%%%%%%%%%%%%%%%%%%%
\subsubsection{Industry}
The ADSP-SC589 from Analog Devices \cite{ADSP} is a processor, consisting of a
standard ARM Cortex-A5 core and two SHARC Digital Signal Processors.
These three processors are connected over a shared memory bus which is
shared with various fixed-function DSP accelerators.
Therefore, this system is very flexible with both the CPU and DSP being fully
independent and programmable, allowing for data to be passed between the CPU
and DSPs in various ways.
However, this device is not designed for radio applications, it is designed for
automotive and entertainment applications.
While Analog Devices do manufacture components designed for radio applications,
they are primarily front-ends and offer little flexibility.

The Qualcomm Hexagon DSP architecture \cite{hexagon}, used in their Snapdragon
processors, are split into two types; the aDSP, used for multimedia applications
and the mDSP, used for the modem.
The Hexagon is a DSP that has been extended to be more general purpose, and
can run a hypervisor that can run a full RTOS or Linux kernel.
This allows significant flexibility, as DSPs usually do not permit running of
high level operating systems.
The Hexagon is connected to the memory bus to communicate with the processor.
This system works ``the other way'' compared to the hybrid system described in
this paper. It adds general purpose execution to a DSP, as opposed to adding
DSP acceleration to a general purpose processor.

%%%%%%%%%%%%%%%%%%%%%%%%%%%%%%%%%%%%%%%%%%%%%%%%%%%%%%%%%%%%%%%%%%%%%%%%%%%%%%%%
\subsection{Further Work}
In this work only a single core CPU was used, so the CPU can only take part in
one contiguous section of signal processing.
So, could multithreading be utilized to replace multiple accelerator blocks,
at different points in the pipeline, or used for multiple parallel processing
pipelines?
Additionally, the accelerator is connected to the CPU via a standard off-chip
I/O bus. So would moving the accelerator closer to the CPU improve the data
movement overhead, reducing latency? For example in this paper~\cite{ringbus}
a neural network accelerator was connected directly to an x86 CPUs ring bus,
improving latency.

In the system described in this paper, when the CPU is woken up by an interrupt
for reads, the cache has to be invalidated before processing can be started.
Could this work be performed before the CPU is woken up?, reducing the time the
CPU is active.
An example of this is Intel's Direct Cache Access technology~\cite{intel1,intel2,intel3}
which allows for a PCIe device to DMA data into the CPU cache, reducing cache
misses, when the CPU is woken up.
Furthermore, cache and interrupt management operations require system calls,
which cause instruction cache misses that stall the CPU.
Can this overhead be reduced by exposing these operations more effectively to
userspace? For example, in this paper~\cite{taskisa} instructions were added to
reduce the communication overhead between the CPU and an external accelerator.
Finally, while the CPU works efficiently during signal processing, it still
takes up a large proportion of time. So could new ISA extensions be used to
reduce the time spent on signal processing?
For example in this paper~\cite{isaext} complex number DSP instructions were added
to a RISC-V processor.

Finally, can this work be applied to other data streaming applications outside
wireless radio that have real-time requirements?

%%%%%%%%%%%%%%%%%%%%%%%%%%%%%%%%%%%%%%%%%%%%%%%%%%%%%%%%%%%%%%%%%%%%%%%%%%%%%%%%
\section*{Acknowledgment}
\addcontentsline{toc}{section}{Acknowledgment}
The authors would like to thank UKRI/EPSRC and Keysight Technologies
for their financial support. UKRI project reference: 2590731

%%%%%%%%%%%%%%%%%%%%%%%%%%%%%%%%%%%%%%%%%%%%%%%%%%%%%%%%%%%%%%%%%%%%%%%%%%%%%%%%
\bibliographystyle{IEEEtran}
\bibliography{IEEEabrv,ref}
\end{document}